\documentclass[5p,authoryear]{elsarticle}
\makeatletter 
\usepackage[utf8]{inputenc}
\makeatletter
\def\ps@pprintTitle{%
  \let\@oddhead\@empty
  \let\@evenhead\@empty
  \def\@oddfoot{\reset@font\hfil\thepage\hfil}
  \let\@evenfoot\@oddfoot
}
\makeatother
\usepackage[T1]{fontenc}
\usepackage[english]{babel}
\usepackage[babel=true]{csquotes}
\usepackage[fleqn]{amsmath} 
\usepackage{amsthm} 
\usepackage{booktabs} 
\usepackage{multirow} 
\usepackage{amssymb} 
\usepackage{float}
\usepackage{hyperref} 
\usepackage[english]{cleveref} 

\begin{document}

\begin{frontmatter}

\title{INCLUSIVE ASTRONOMY IN PERU: CONTRIBUTION OF ASTRONOMY TEACHING FOR VISUALLY IMPAIRED PEOPLE} 

\author{\large Alexis Rodriguez Quiroz\textsuperscript{1}, \large Kevin Vidal Céspedes\textsuperscript{1}, \large  María Argudo-Fernández\textsuperscript{2} \\ \bigskip \normalsize \textit{1} \textit {Universidad Nacional Mayor de San Marcos} \textit {(UNMSM) \\ \normalsize \textit{2} \textit {Instituto de Física, Pontificia Universidad Católica de Valparaíso} \textit {(PUCV)}}}

\begin{abstract} 

Everything we know about the environment around us is thanks to light, a kind of electromagnetic radiation. Astronomy takes advantage of it and all the electromagnetic spectrum with the help of many devices to record them and determine from which places in our Universe they come. These signals must be processed to obtain the images that are will be then exposed to the public. Astronomers know that these images will inspire and generate curiosity in each person who sees them. This Science is inclusive and wants to transmit the knowledge and the beautiful events that happen in the universe to all people. We try to do that by developing and showing other forms of teaching, taking advantage of new technologies that are available today to bring this knowledge closer to the minorities in our country like visually impaired people.

Because of this problem, the AstroBVI project arrived in Peru for the first time, thanks to the distribution of tactile images of 3D galaxies that were delivered to us from the Centro de Astronomía de la Universidad de Antofagasta of the and financed by the International Astronomical Union Office of Astronomy for Development (IAU-OAD). This allowed the holding of seven workshops during 2019, visiting various institutions and benefiting more than 160 participants with blindness and low vision, identifying in them a lot of interest, which shows the enormous potential presented by these 3D tactile materials. \\

\end{abstract}


\end{frontmatter}

\section{Introduction}\label{introduction}
According with the last national census organized by Instituto Nacional de Estadística e Informática, \cite{perfil2017}, most prevalent disability in Peru is the visual disability (48.3\%) which represent 1 473 583 (573 389 males and 900 194 women), concentrating the largest amount in Lima (564 252 people), which represents 38.29\%.

Related with the education of this group of people, the Sociodemographic profile, \cite{pobdiscapacidad2017} shows that 13\% of the total of persons over 3 years (3 189 379) with some disabilities never had the opportunity to get access to the educational level in their lives, just the 0,5\% (17 232) had the privilege to get access at a study center. All of these indicators reveal the importance of attending and helping this minority, that in many cases are excluded, affecting their intellectual development and making so hard their integration to the peruvian society.

The strategic plan 2010-2020 of the International Astronomical Union (IAU)\footnote{The IAU Strategic Plan 2010 – 2020: Astronomy for the \\Developing: https://www.iau.org/education/strategic\_plan/} shows the importance of Astronomy as a tool to promote technology, science and cultural development. In this plan, it mentions how Astronomy can contribute to Junior School and High School; it is through this that AstroBVI\footnote{AstroBVI: https://astrobvi.org} is gestated as an educational inclusion project that was designed for the community made up of people with blindness and visual impairment in Latin America, which responds to the acronym in English BVI (blind and visually impaired) so that, through teachers, they can acquire basic and intermediate knowledge of Astronomy and at the same time raise awareness in the community and among the general public.

Testimonies of the participants in our workshops, including adults and high school students, informed us that among the classes received in Science they briefly touch on topics about the Earth in the Solar System and its internal processes, this can be corroborated in the National Curriculum of Basic Education, Science and Technology area, in the competence: “Explains the physical world based on knowledge about living beings, matter and energy, biodiversity, Earth and the Universe”, both for Junior School \cite{promprimaria} and High School \cite{promsecundaria}, whose general curriculum by the Ministry of Education is adapted or incorporates capacities to respond to students with special educational needs.

This gives us an indication that the approach to Science for the visually impaired public is very few due to social, pedagogical and material limitations that are scarce in our country.

We carried out this experience between April and November 2019, for a total of seven sessions. This article is to show the situation of basic special education, in the case of Astronomy, in Peru and tell our experience when we gave the workshops to the public that was very interested in listening and learning new and fascinating things.

\section{Background of Inclusive Astronomy}
Inclusive astronomy has grown in the last 20 years, showing a variety of alternatives to bring this Science to visually impaired people. We understand by inclusion, the fact that Astronomy is capable of making possible the participation and integration of students and the general public, regardless of their physical characteristics, social and intellectual; in scientific matters and generate equality in the acquisition of knowledge. 

One of the strategies to teach Astronomy to the visually impaired community is the use of high-relief images  to represent craters,  characteristics and intensity of light of the objects of the universe. Its sensory capacity allows them to recognize and receive information about an object.

There are many examples of these images compiled in books. Two of the first books published on relief and Braille was “Touch the Universe: A NASA Braille Book of Astronomy” (Noreen Grice, 2002) and “Touch the Invisible Sky” (Noreen Grice, Simon Steel \& Doris Daou, 2007), however they cannot detail the complexity of the structures that the different celestial bodies present.

Due to this, more proposals were presented such as 3D printed tactile images, such as those executed by The Tactile Universe project\footnote{The Tactile Universe: https://tactileuniverse.org} carried out by the Institute of Cosmology and Gravitation at the University of Portsmouth, United Kingdom, this project used printed 3D images about galaxies. That project has materials for teachers which can be downloaded for free on the website. Another project was A Touch of the Universe project\footnote{A Touch of the Universe: https://astrokit.uv.es} of the Astronomical Observatory by the University of Valencia, which presents tactile materials on planets, the moon and the constellations (The sky in your hands). 

In addition to tactile materials, auditory material is being used as a complement in teaching, such as audiobooks and collections of files with sounds from space by NASA, these are obtained by capturing electromagnetic wave emissions, and then converting them into sound waves. 

One of the initiatives to highlight is the device called Ligthsound\footnote{Ligthsound: http://astrolab.fas.harvard.edu/LightSound-\\IAU100.html}, created by Allyson Bieryla, Rob Hart and Daniel Davis of Harvard University in collaboration with Wanda Diaz of IAU Office of Astronomy of Development. The device was used in the eclipses of 2017 at United States, at Chile in 2019 and 2020, which allowed the visually impaired public to experience this phenomenon. It detects the change in brightness and converts the light into sound, in such a manner that the sound will be a higher pitch when there is more light and a lower pitch when the Sun is being dwarfed by the Moon (Allyson Bieryla, 2017).

\subsection{\textbf {Inclusive Astronomy in South America}}
This initiative is growing in South America, we can mention   educational projects such as Astronomy with all the senses of the Medellín Planetarium\footnote{Medellín Planetarium: https://planetariomedellin.org} in Colombia, where they present a tactile material on constellations, the Sun, the Moon, among other celestial bodies. Also at the Galileo Galilei Planetarium\footnote{ Galileo Galilei Planetarium: https://planetario.buenos \\aires.gob.ar}, in Argentina, they present functions called “Heaven for all”, where they use tactile and auditory material. In Chile there are some dissemination projects using tactile material such as those caried out by the Dedoscopio group\footnote{Dedoscopio: https://www.facebook.com/dedoscopio/}, Nucleo de Astronomía UDP group\footnote{Nucleo de Astronomía UDP: https://astronomia.udp.cl/}  and AstroBVI, which also together with other initiatives have joined at the national level creating an Inclusive Astronomy group\footnote{Grupo Astronomía Inclusiva: https://www.facebook.com/ \\ astroinclusiva}.

\subsection {\textbf {The AstroBVI project}}
AstroBVI is one of the most important projects in Chile, made possible thanks to funding in 2018 by the Office of Astronomy for Development of the International Astronomical Union (IAU- OAD) which was one of the 16 projects selected by the sixth annual call of proposals, within the Task Force 2: Astronomy for Schools and Children\footnote{Results of the IAU-Office of Astronomy for Development’s 2017 https://www.iau.org/news/announcements/detail/ann18003/}. This initiative is based on the foundations of projects mentioned above, especially the tactile images of 3D galaxies that were developed in consultation with The Tactile Universe. Thanks to this project, many countries like (Argentina, Bolivia, Brazil, Chile, Colombia, Costa Rica, Ecuador, Honduras, Mexico, Paraguay, Peru and Venezuela) but also in Spain and Nepal, \cite{capjournal}, were benefited with the materials to be able to carried out educational activities in schools and the general public. Thanks to this project, we were able to introduce Inclusive astronomy in Peru.

\section{Didactic material}
One of the most developed senses by the visually impaired public is touch. With it they can recognize and identify a diversity of features of an object. In order to explain and develop our class, we turned to 3D printed relief images.
The relief of these materials was what allowed us to convey the idea of how celestial objects are like and what form they have. This outreach strategy, in addition to being the most inclusive from the point of view of the visually impaired people, is the simplest and requires the least number of additional means, \cite{astroaccesible2017}.

The main element we used in the workshops was the educational kit provided by AstroBVI, which is made up of the 3D tactile images of different galaxies and at different wavelengths, \cite{argudo2019astrobvi}.

\begin{itemize}
\item M100 is a spiral galaxy without a bar, the image of which was captured in the photometric B band.
\item M109 is an inclined barred spiral galaxy.
\item M51 is a fusion between a spiral and elliptical galaxy, captured in the R and B bands.
\item M105 is an elliptical galaxy observed in the R band.
\item NGC5866 is a lenticular galaxy captured in the R band.
\item Digital images that help people with low vision to compare and have a reference of the 3D tactile images.
\end{itemize}

\begin{figure} [h]
    \centering
	\centerline{\includegraphics[width=3.4in]{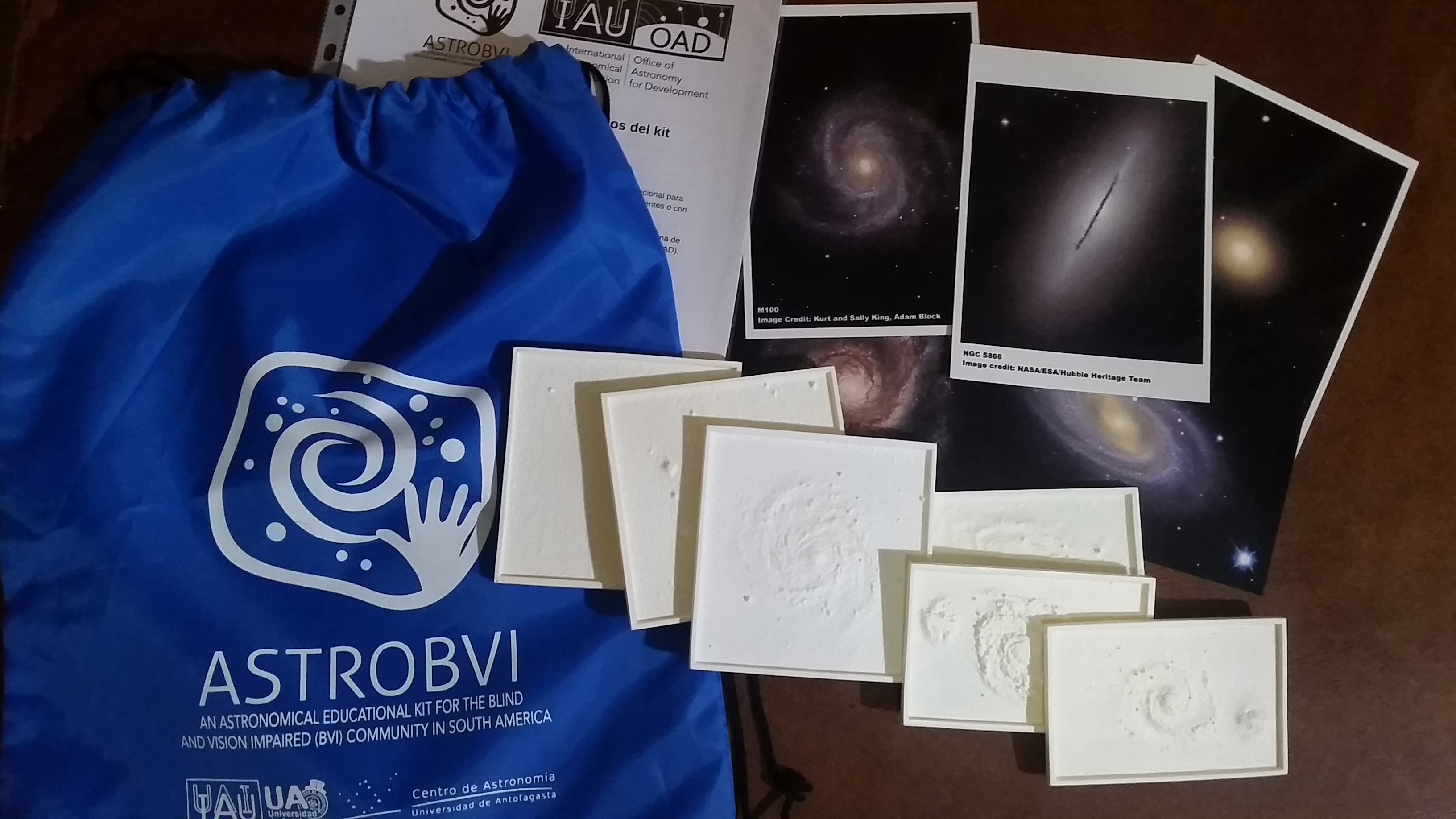}}
	\caption[]{Relief images of the galaxies by AstroBVI project. Own photography.}
	\label{fig}
\end{figure}

In addition to these materials, we used the following elements as a complement, to help generate a broader vision of astronomical objects:

\begin{itemize}
\item The 3D printed Moon allows participants to know in a good manner about the craters and the surface that exist in it.
\item The Solar System, made of cold ceramics, which gave an idea of the sizes of the planets and order. Toy balls of different sizes were also used.
\item Auditory material, which was downloaded from the NASA library’s spatial sound files to be used to help students know the interior and the phenomena that occurred on each planet\footnote{ NASA: Spooky Sounds from Across the Solar System: https://soundcloud.com/nasa/sets/spookyspacesounds}.

\end{itemize}

\begin{figure} [h]
    \centering
	\centerline{\includegraphics[width=3.4in]{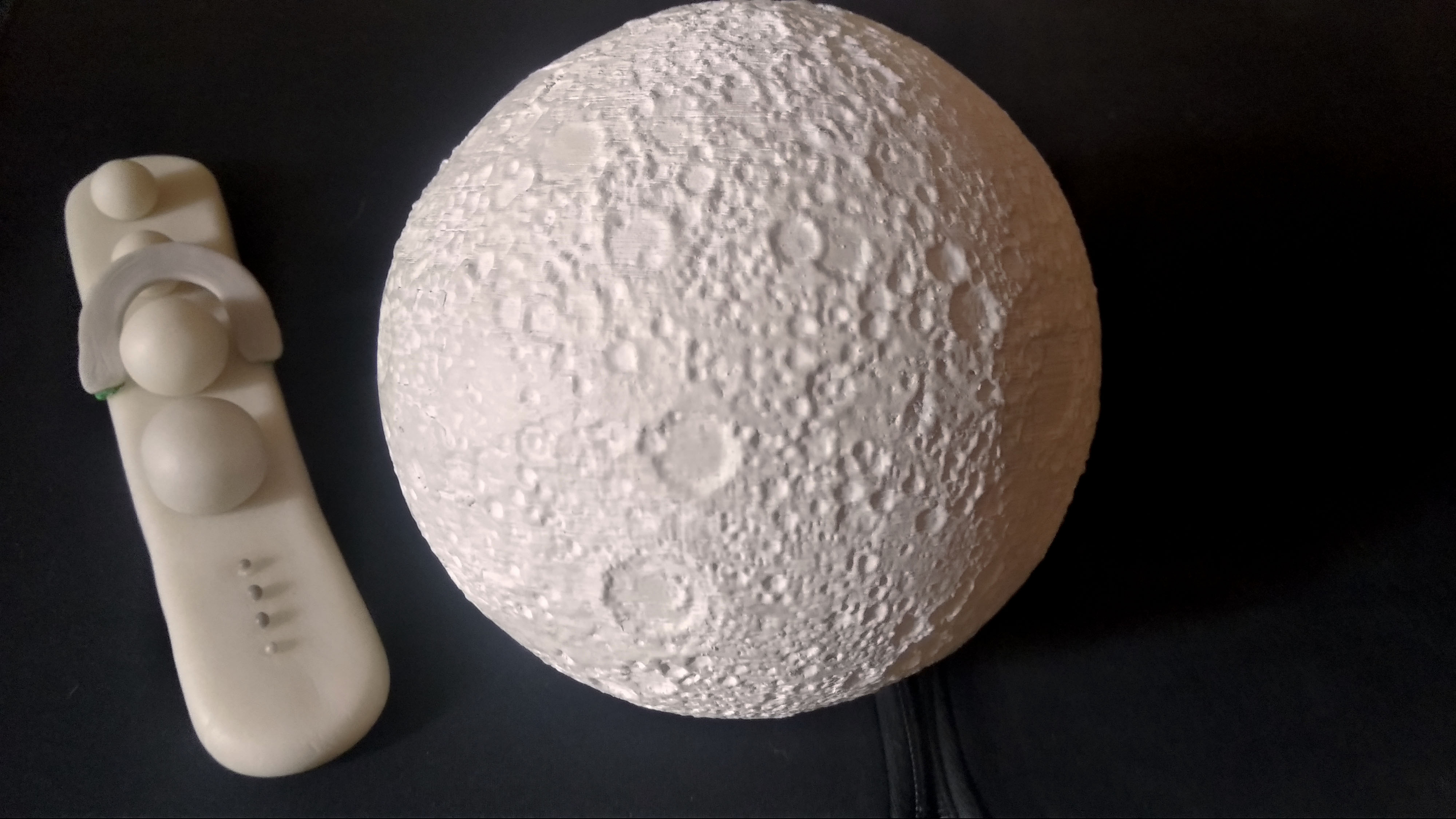}}
	\caption[]{3D printed relief and surface of the Moon and representations of the planets. Own elaboration.}
	\label{fig}
\end{figure}

\section{Methodology}
For each workshop we propose a strategy to carry out the best possible experience and understanding for the participants. In each step of the development, the limitations of the activity were taken into account, considering that many of the participants have no previous knowledge of Astronomy.

\begin{figure} [h]
    \centering
	\centerline{\includegraphics[width=2.98in]{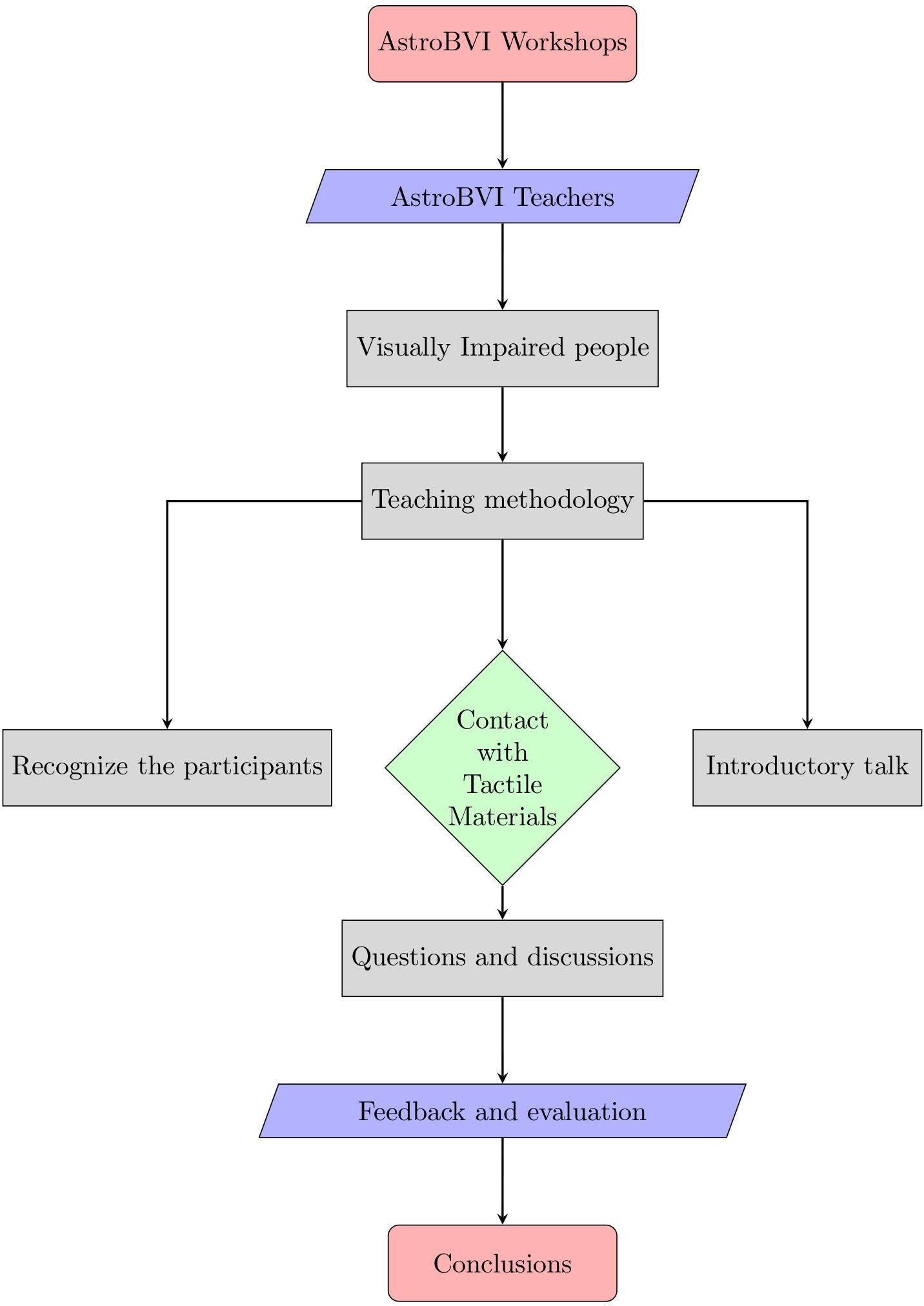}}
	\caption[]{Teaching methodology diagram. Own elaboration.}
	\label{fig}
\end{figure}

That is why, at the beginning of each session, we recognize the participants, this means we ask them to give us some ideas of a concept on the subject to be discussed, it is where you can get the points to reinforce and how to manage the session. After that, an introductory talk is held in the simplest and most didactic way, so that later they can consolidate what they have learned using the tactile materials. In some cases, the auditory material helped the participants discover more about the celestial bodies.

In each workshop it was explained the concepts of galaxies, formation, electromagnetic spectrum and components of these and what information represents the reliefs of the plaques. On the Moon, it was explained about its formation, the surface and the importance of it to Earth and also about the Solar System, the planets, sizes and the Sun. 

This generated questions and discussions that occurred when making contact with the material; many people were surprised and for some it was the first time they knew these celestial bodies, as they were small groups, so it facilitated a closer interaction with them, so that we could evaluate their acquired knowledge by asking them questions and a little feedback.

Finally, they proceeded to draw conclusions from the topics discussed and on the experience of the workshop in general.

\section{Astronomy teaching workshops}
Each workshop had a different strategy, this depended on the target audience, infrastructure, and the coordinators of each institution. Seven sessions were held in total, where we can see how each one developed: 

Our first workshop was held in April, in the Unión de Ciegos del Perú (UDCP), a place that brings together a large group of young people and adults between 23 and 72 years old, with visual limitation problems. This was held on April 10th with an attendance of 11 participants.

\begin{figure} [h]
    \centering
	\centerline{\includegraphics[width=3.4in]{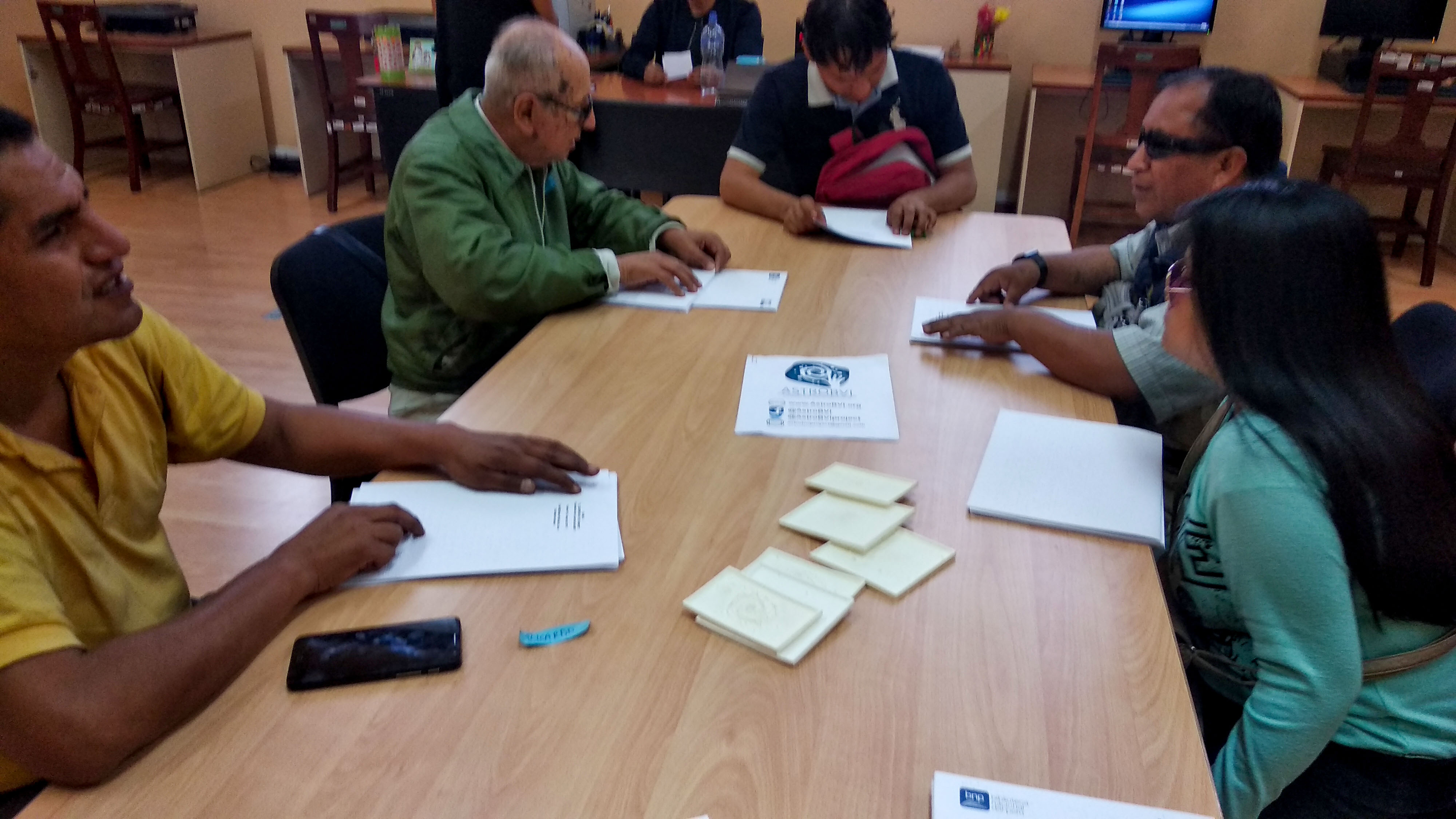}}
	\caption[]{Activity carried out in the room for the blind of the Great Public Library of Lima. Own photography.}
	\label{fig}
\end{figure}

The second workshop was on May 17th at the “Delfina Otero Villarán” Room for the blind by Gran Biblioteca Pública de Lima, this time they supported us with material in Braille System with additional information on the topics discussed, we had an audience between 15 to 75 years old people, with an attendance of 13 participants.

We also visited special educational institutions, such as the Special Basic Education Center (C.E.B.E). Luis Braille High School on July 17th and we had a special date for the 50th Anniversary of the Moon Landing, held on July 24th. On both days there were 31 participants between 13 and 27 years old. Subsequently, we held two workshops at the C.E.B.E San Francisco de Asís for Kindergarden and Junior School on September 23rd and 25th, where we included auditory material since they had a multimedia room. On both days we had 99 participants between 6 to 19 years old and we culminated with a second date at the Gran Biblioteca Pública de Lima on November 15th with 11 participants between 15 to 78 years old.

\begin{figure} [h]
    \centering
	\centerline{\includegraphics[width=3.4in]{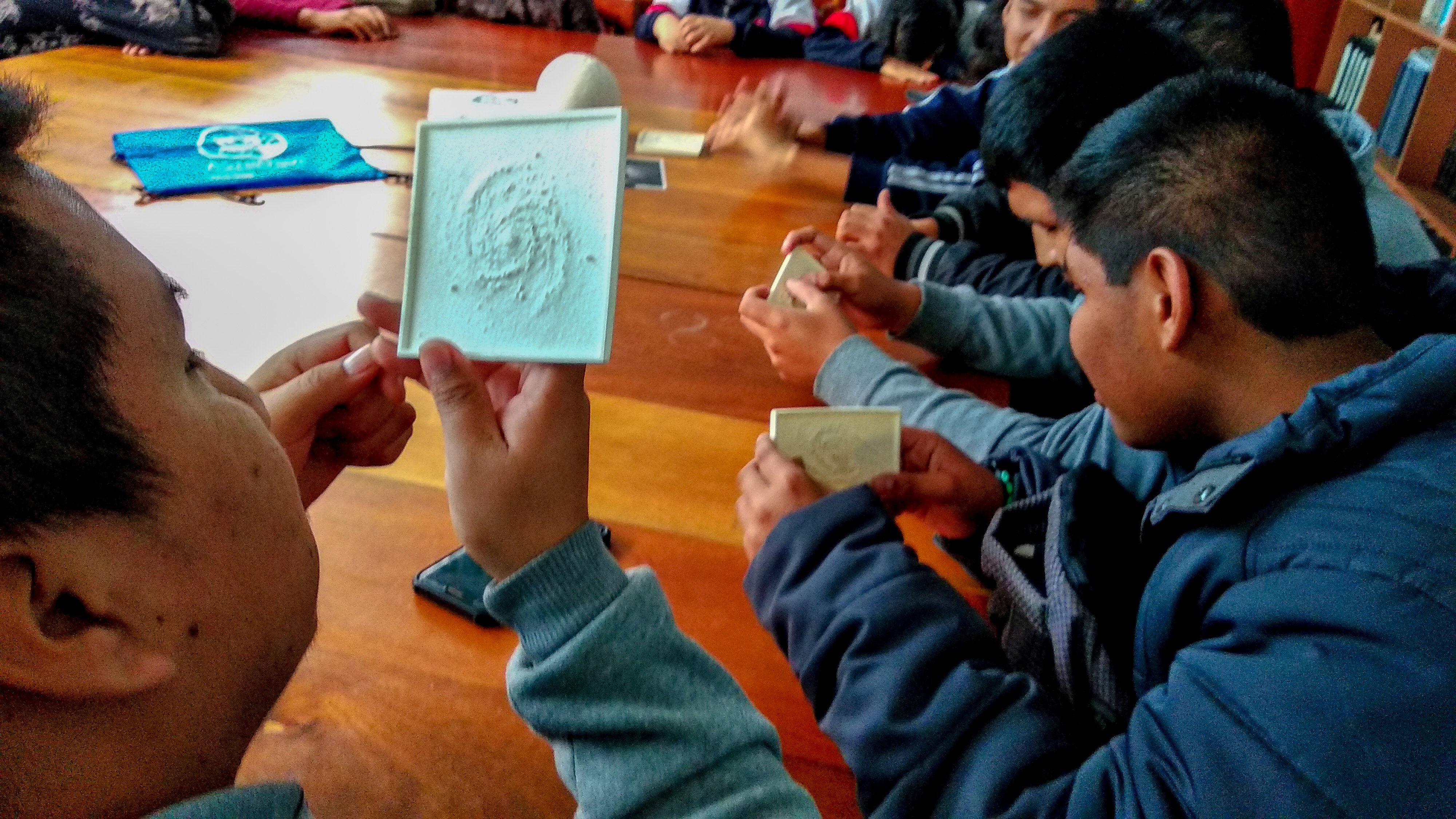}}
	\caption[]{Activity carried out at the C.E.B.E. Luis Braille. Own photography.}
	\label{fig}
\end{figure}

\section{Impact of the project}
The AstroBVI project has had a positive impact because it managed to reach the target audience, who are BVI people. In total, we had 165 BVI participants in seven workshops, where two groups can be distinguished whose visual limitations range from low vision (68 people) to blindness (97 people). The ages of the participants in these workshops were mostly under 20 years old, since the project could be taken to schools and to a lesser extent there were people between 20 and 75 years old. 

In this way, we were able to participate in the activities of The International Astronomy Day\footnote{The International Astronomy Day: \\ https://blog.astrobvi.org/post/184345894577/}, it was also possible to visit important institutions, where until today this type of activity had not been carried out.

The project was disseminated in various media, in particular, a TV show dedicated to people with disabilities called Sin Barreras\footnote{Sin Barreras TV Show: https://www.tvperu.gob.pe/novedades/sin-barreras/conoce-el-proyecto-que-acerca-la-astronomia-a-personas-con-discapacidad-visual}, where we were able to show the potential it offers and the impact that the AstroBVI project has generated.

We were accepted by the Peruvian Physical Society to be able to present ourselves at the XXVIII Peruvian Physics Symposium\footnote{XXVIII Peruvian Physics Symposium: \\ http://www.soperfi.org/web/spf2019/orales.pdf}, held by the University of Trujillo, to present the results of teaching Astronomy to the visually impaired public. Here we show our results:

\begin{figure}[h]
    \centering
	\centerline{\includegraphics[width=3.6in]{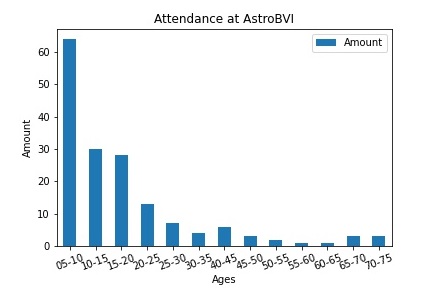}}
	\caption[]{Total number of attendees with a degree of disability. A total of 68 people with low vision and 97 with blindness are shown.}
	\label{fig}
\end{figure}

\begin{figure}[h]
    \centering
	\centerline{\includegraphics[width=3.6in]{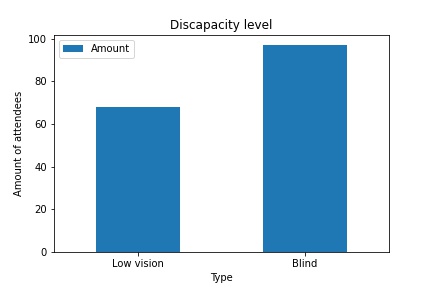}}
	\caption[]{Total number of ages of those attending the seven workshops held. A total of 122 people under 20 years of age are shown, 24 people between 20 to 35, 11 people between 40 to 55 and 8 people over 55 years old.}
	\label{fig}
\end{figure}

\begin{figure}[h]
    \centering
	\centerline{\includegraphics[width=3.6in]{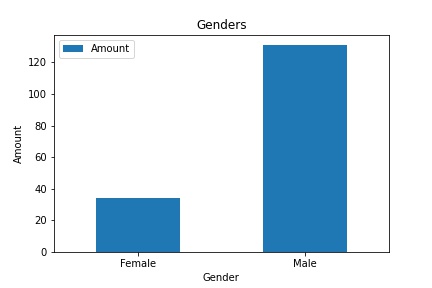}}
	\caption[]{Total gender of attendees in the seven workshops. 34 attendees of the female audience are shown, which becomes 21\% of the attendance and a total of 131 male attendees.}
	\label{fig}
\end{figure}

\section{Conclusions}
The AstroBVI project deployed at Lima city, showed great interest of the visually impaired community to learn about Science. This was proven by the attendance, curiosity and disposition of the public about the project and the topics presented, as well as by the institutions that helped in the development of the workshops.

We can conclude that the teaching methodology has been effective, since which evidenced the learning of the participants by responding positively to each stimulus at the end of each workshop, due to the formula between introductory talks, questions and consolidation with the tactile material. 

The use of suitable materials, knowledge and adequate training on special education by the teachers, allowed them to reach people who were not involved with Astronomy, breaking down barriers between Science and visual impairments.

These workshops made it possible to learn about the problems faced by visually impaired people when learning about science-related topics, in this case about Astronomy. There is a great lack of material for these people and too little help. 

\section{Acknowledgements}
We want to express our gratitude to the group involved in the AstroBVI project by Centro de Astronomía de la Universidad de Antofagasta who provided us with the materials with which we were able to make the workshops possible. To the physicist Ricardo Ángelo Quispe Mendizábal and to the teacher Fernando Camacho Zelaya for their help in the workshops. 

Also to the C.E.B.E. Luis Braille, San Francisco de Asís, la Gran Biblioteca Pública de Lima, la Unión de Ciegos del Perú (UDCP), el Instituto Geofísico del Perú (IGP), la Sociedad Peruana de Física (SOPERFI) and the media who allowed us to show the project.

\bibliography{bibliography.bib}

\end{document}